\documentclass[aps,prc,twocolumn,nofootinbib,amsmath,showpacs,superscriptaddress,floatfix]{revtex4}
\usepackage{graphicx,epsfig,longtable}
\usepackage{mathptmx}
\usepackage[light,first]{draftcopy} 
\usepackage{epstopdf}
\usepackage[pdftex]{hyperref}
\usepackage{wasysym}
\usepackage{color}

\newcommand{\beqy}{\begin{eqnarray}}
\newcommand{\eeqy}{\end{eqnarray}}
\newcommand{\bmlet}{\begin{subequations}}
\newcommand{\emlet}{\end{subequations}}
\newcounter{saveeqn}

\def\gsimeq{\,\,\raise0.14em\hbox{$>$}\kern-0.76em\lower0.28em\hbox  
{$\sim$}\,\,}  
\def\lsimeq{\,\,\raise0.14em\hbox{$<$}\kern-0.76em\lower0.28em\hbox  
{$\sim$}\,\,}  

\begin{document}

\title{Validity of the generalized Brink-Axel hypothesis in $^{238}$Np}

\author{M.~Guttormsen}
\email{magne.guttormsen@fys.uio.no}
\affiliation{Department of Physics, University of Oslo, N-0316 Oslo, Norway}
\author{A.~C.~Larsen}
\email{a.c.larsen@fys.uio.no}
\affiliation{Department of Physics, University of Oslo, N-0316 Oslo, Norway}
\author{A.~G{\"o}rgen}
\affiliation{Department of Physics, University of Oslo, N-0316 Oslo, Norway}
\author{T.~Renstr{\o}m}
\affiliation{Department of Physics, University of Oslo, N-0316 Oslo, Norway}
\author{S.~Siem}
\affiliation{Department of Physics, University of Oslo, N-0316 Oslo, Norway}
\author{T.~G.~Tornyi}
\affiliation{Department of Physics, University of Oslo, N-0316 Oslo, Norway}
\author{G.~M.~Tveten}
\affiliation{Department of Physics, University of Oslo, N-0316 Oslo, Norway}

\date{\today}

\begin{abstract}
We have analyzed primary $\gamma$-ray spectra of the odd-odd $^{238}$Np nucleus extracted 
from $^{237}$Np($d,p\gamma$)$^{238}$Np coincidence data measured at the Oslo Cyclotron Laboratory.
The primary $\gamma$ spectra cover an excitation-energy region of $0 \leq E_i \leq 5.4$ MeV,
and allowed us to perform a detailed study of the $\gamma$-ray strength as function of 
excitation energy. Hence, we could test the validity of the generalized Brink-Axel hypothesis,
which, in its strictest form, claims no excitation-energy dependence on the $\gamma$ strength.
In this work, using the available high-quality $^{238}$Np data, 
we show that the $\gamma$-ray strength function is to a very large extent independent on
the initial and final states.
Thus, for the first time, the generalized Brink-Axel hypothesis
has been experimentally verified for $\gamma$ transitions between states in the quasi-continuum region,
not only for specific collective resonances, but also for the full strength below the neutron
separation energy. Based on our findings, the necessary criteria for the generalized
Brink-Axel hypothesis to be fulfilled are outlined.
\end{abstract}


\maketitle

Sixty years ago, David~M.~Brink proposed in his PhD thesis~\cite{brink} that
the photoabsorption cross section of the giant electric dipole resonance (GDR)
 is independent of the detailed structure of the initial state.
In his thesis, he expressed his hypothesis as follows: {\em  "If it
were possible to perform the photo effect on an excited state, the cross section
for absorption of a photon of energy $E$ would still
have an energy dependence given by (15)"}, where equation (15) refers
to a Lorentzian shape of the photoabsorption cross section.
Brink's original idea, the \textit{Brink hypothesis}, was first
intended for the photoabsorption process on the GDR, but
has been further generalized, applying the principle of detailed balance,
to include absorption and emission of $\gamma$ rays between resonant states~\cite{axel,Bartholomew}.
In addition to assuming independence of excitation energy, 
there is no explicit dependence of initial and final spins except the obvious
dipole selection rules, implying that all levels exhibit the same dipole strength
regardless of their initial spin quantum number.
We will refer to this as the \textit{generalized Brink-Axel (gBA) hypothesis}.
A review of the history of the hypothesis was given 
by Brink in Ref.~\cite{oslo2009}.

The gBA hypothesis has implications for almost
any situation where nuclei are brought to an excited state above $\approx 2\Delta$, 
where $\Delta \approx 1$ MeV is the pair-gap parameter. 
Here, the nucleus will typically de-excite via $\gamma$-ray emission
and/or by emission of particles. In this context, it is usual to translate the
$\gamma$-ray cross section $\sigma(E_{\gamma})$ into $\gamma$-ray strength function ($\gamma$SF)
by $f(E_{\gamma})=(3\pi^2\hbar^2c^2)^{-1}\sigma(E_{\gamma})/E_{\gamma}$.

To describe and model the electric dipole part of the $\gamma$-decay channel, the gBA hypothesis is
frequently used, applying in particular the assumption of spin independence~\cite{RIPL3}. 
For example, a rather standard approach to calculating $E1$ strength is to apply some variant of the
quasi-particle random-phase approximation (QRPA) to obtain $B(E1)$ values as function of excitation energy,
and assuming that this $E1$ distribution corresponds to the one in the quasi-continuum;
see, e.g., Ref.~\cite{daoutidis2012} and references therein.
Also for $M1$ transitions the gBA hypothesis has been utilized, see e.g. Ref.~\cite{loens2012}.
Furthermore, the hypothesis is also often applied to $\beta$-decay and electron capture
for calculating Gamow-Teller and Fermi transition strengths, 
see e.g. Ref.~\cite{fantina2012} and references therein. The main reason for its wide
range of applications is the drastic simplification of the considered problem, and
in some cases it is a key necessity to be able to perform the desired calculation~\cite{fantina2012}.
Hence, the question of whether the hypothesis is valid or not, and under which circumstances,
is of utmost importance for multiple reasons: its fundamental impact on nuclear structure
and dynamics, and its pivotal role for the description of $\gamma$ and $\beta$ decay
for applied nuclear physics, such as input for ($n,\gamma$) cross-section calculations
relevant to the $r$-process nucleosynthesis in extreme astrophysical
environments~\cite{arnould2007} and next-generation nuclear power plants~\cite{IAEA-TECDOC}.

However, it is not at all obvious neither from 
experiment nor theory that the gBA hypothesis is valid.
From an experimental point of view, there are two main reasons for this;
the hypothesis has primarily been tested at very high excitation energies or with
only a few states included.
In the first case, compilations show that the width of the GDR varies
with temperature and spin, in contradiction to the gBA hypothesis~\cite{schiller2007}. However,
the hypothesis was not originally considered for building the GDR on such highly excited states.
Obviously, thermal fluctuations will affect the width of the GDR,
but the GDR energy centroid stays rather fixed.
Other test cases suffer from large Porter-Thomas (PT) fluctuations~\cite{PT}, since
the $\gamma$SF could not be averaged over a sufficient amount of levels~\cite{46Ti}.
In particular, this is the case for lighter nuclei or if levels close to the ground state are considered.

In general, experimental data supporting the gBA hypothesis are rather scarce.
For example, $(n,\gamma)$ reactions give
$\gamma$SFs consistent with the gBA hypothesis, but in a limited
$\gamma$-ray energy range~\cite{stefanon1977,raman1981,kahane1984,islam1991,kopecky1990}.
Furthermore, data on the $^{89}$Y$(p,\gamma)^{90}$Zr
reaction point towards deviations from the gBA hypothesis~\cite{netterdon2015}.
There have also been various theoretical attempts to test the gBA hypothesis and modifications or
even violations are found~\cite{misch2014,johnson2015}.
For some theoretical applications, the assumption
of the gBA hypothesis is successfully applied~\cite{koeling1978,horing1992,gu2001,betak2001,hussein2004}.
We may learn from these experimental and theoretical attempts that
the structure and dynamics of the system represents important constraints.

In this Letter, we address the gBA hypothesis
from an experimental point of view, and we provide the needed criteria for the hypothesis
to be valid for $\gamma$ decay below the neutron threshold by a detailed analysis of the
$^{238}$Np $\gamma$SF.
The $^{238}$Np nucleus is probably the ultimate case to test the gBA hypothesis,
as it is an odd-odd system with extremely high level density.
Already a few hundred keV above the ground state, we find a level density of $\approx 200$~MeV$^{-1}$,
which increases to $\approx 43\cdot 10^{6}$~MeV$^{-1}$ at the neutron separation energy of $S_n=5.488$~MeV.
In a previous study~\cite{238Np}, the level density and $\gamma$SF were
extracted from the distributions of primary $\gamma$-rays measured in the
$^{237}$Np$(d, p \gamma)^{238}$Np reaction. This very rich data set represents ideal
conditions for testing the gBA hypothesis where
the PT fluctuations are negligible due to the high level density.
In the following, we utilize the primary matrix of initial excitation energy $E_i$ versus $\gamma$-ray
energy~\cite{238Np}.

\begin{figure}[t]
\begin{center}
\includegraphics[clip,width=0.85\columnwidth]{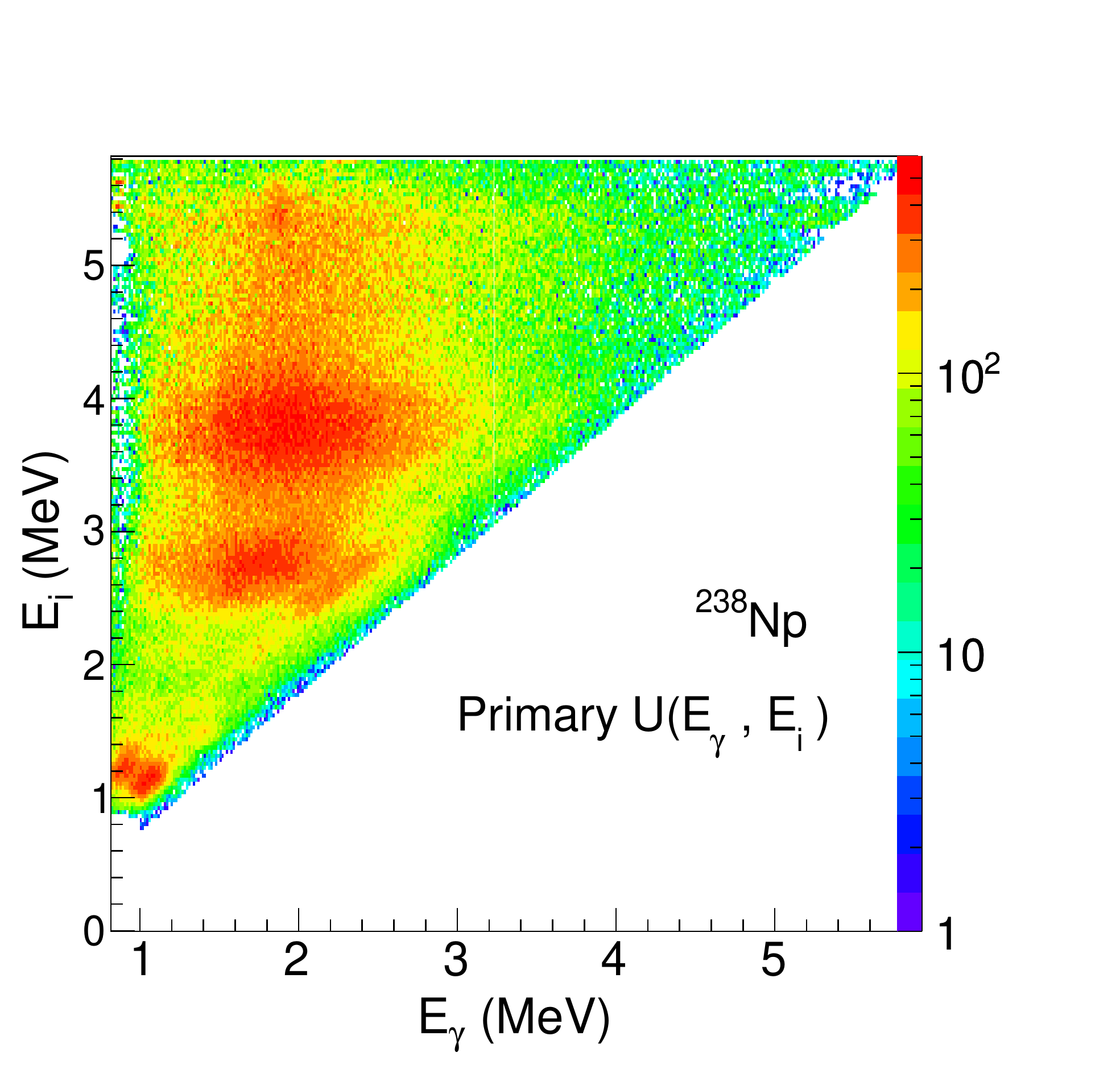}
\caption{(Color online) Primary $\gamma$-ray matrix of $^{238}$Np~\cite{238Np}.}
\label{fig:matrix}
\end{center}
\end{figure}
\begin{figure}[t]
\begin{center}
\includegraphics[clip,width=1.02\columnwidth]{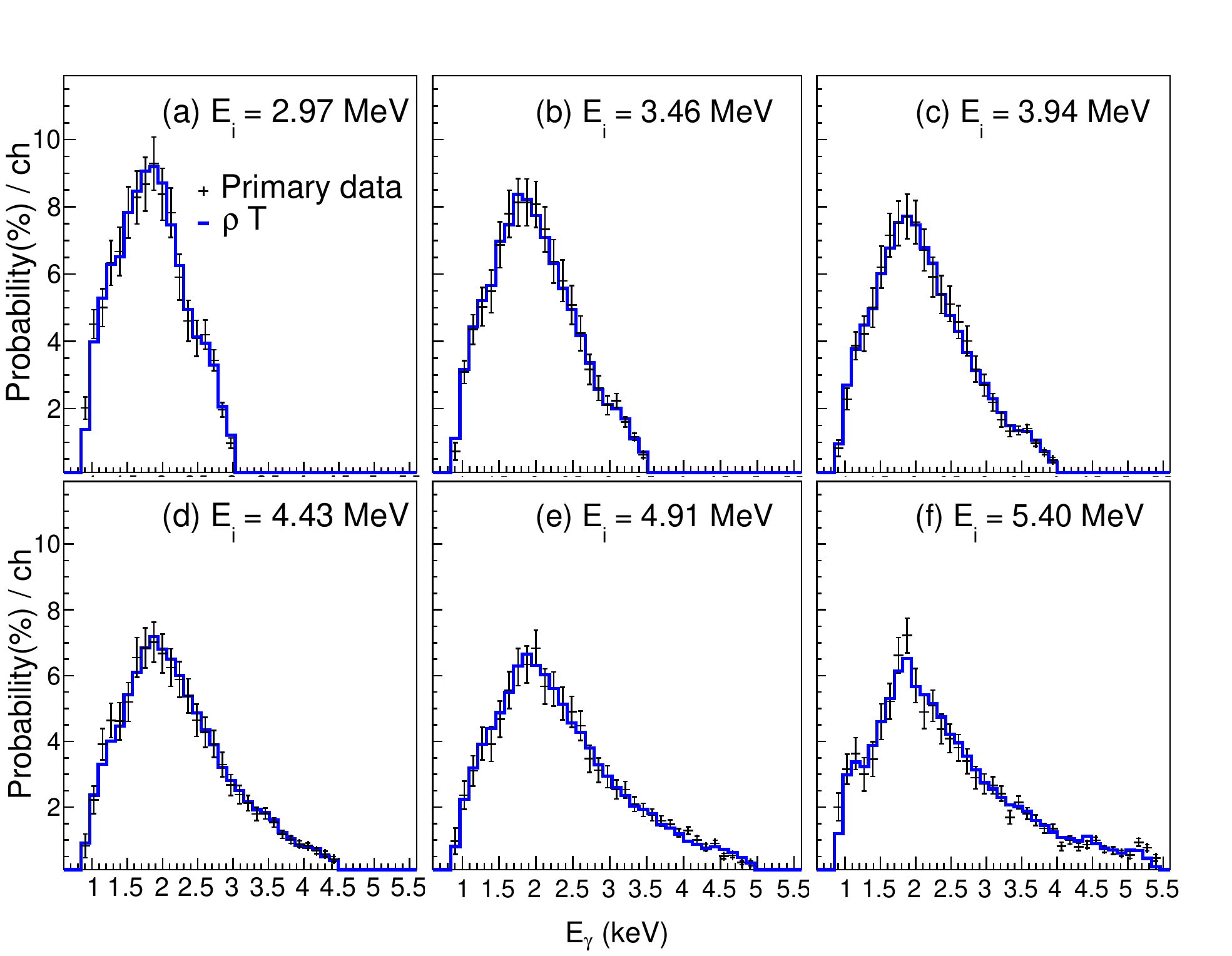}
\caption{(Color online) Primary $\gamma$-ray spectra (data points) at various initial excitation energies
compared to the product $\rho(E_i-E_{\gamma}){\cal{T}}(E_{\gamma})$ (blue curve).}
\label{fig:doesitfit}
\end{center}
\end{figure}
\begin{figure}[]
\begin{center}
\includegraphics[clip,width=0.8\columnwidth]{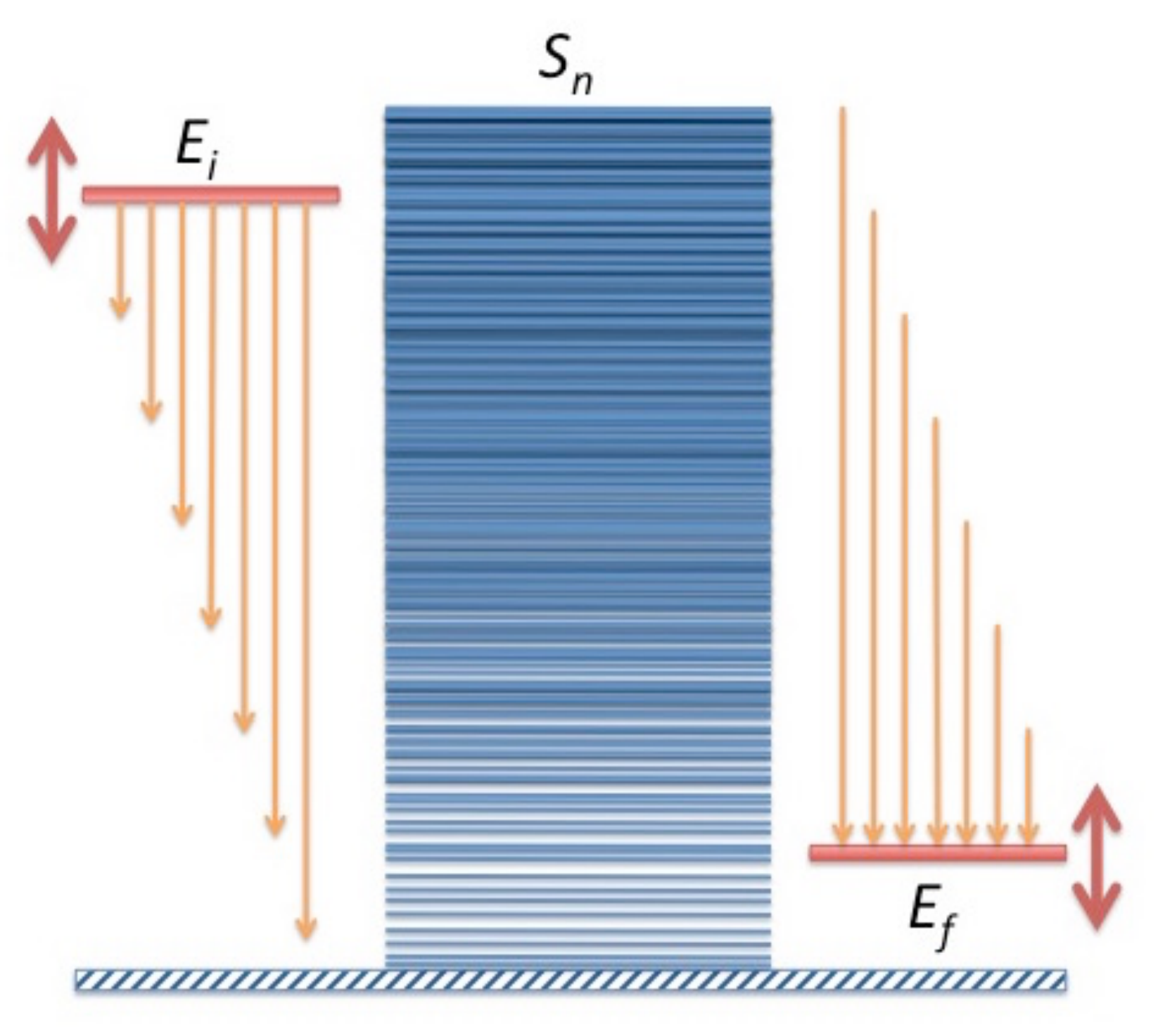}
\caption{(Color online) The procedure to extract $\gamma$SFs as function of
initial $E_i$ (left) and final $E_f$ (right) excitation energies.
The blue-shaded region (middle) illustrates the exponentially increasing
level density as function of excitation energy. The two $\gamma$SFs
are limited to $E_{\gamma} < E_i$ and
$E_{\gamma} < S_n - E_f$, respectively, where $S_n$ is the neutron separation energy.}
\label{fig:method}
\end{center}
\end{figure}
\begin{figure}[]
\begin{center}
\includegraphics[clip,width=0.9\columnwidth]{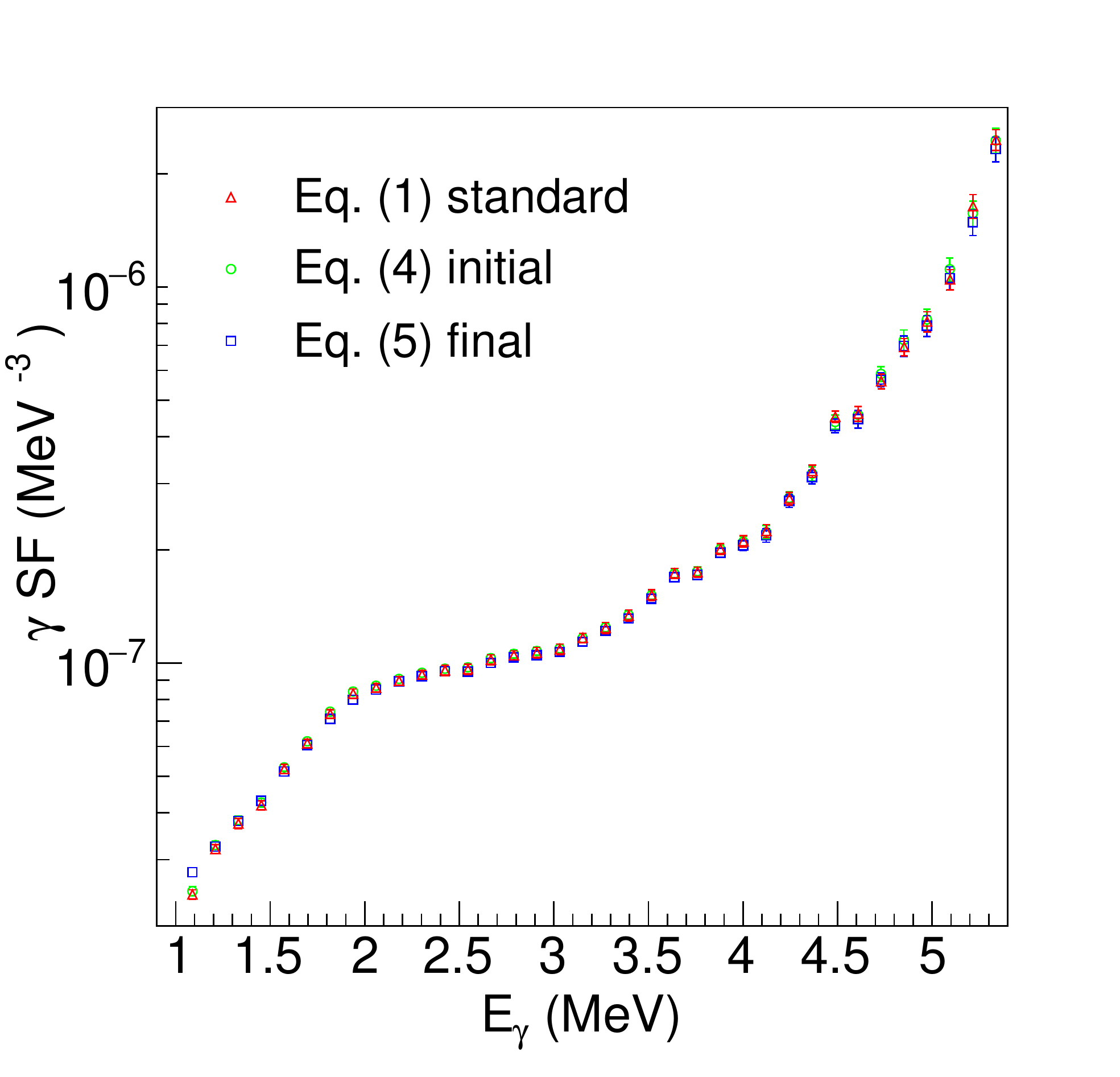}
\caption{(Color online) Comparison between the three $\gamma$SFs obtained by
Eqs.~(\ref{eq:rhoT}), (\ref{eq:Ei}) and  (\ref{eq:Ef}).}
\label{fig:strength3}
\end{center}
\end{figure}

Figure~\ref{fig:matrix} shows the primary $U(E_{\gamma},E_i)$ $\gamma$ spectra
(unfolded with the detector response functions)
as function of initial excitation energy $E_i$.
We now normalize $U$ to obtain the probability that the nucleus emits
a $\gamma$-ray with energy $E_{\gamma}$ at excitation energy $E_i$ by
$P(E_{\gamma},E_i )= U(E_{\gamma},E_i)/\sum_{E_{\gamma}}U(E_{\gamma},E_i)$.
The probability is assumed to be factorized into:
\begin{equation}
P(E_{\gamma},E_i ) \propto   \rho(E_i-E_{\gamma}){\cal{T}}(E_{\gamma}) .\
\label{eq:rhoT}
\end{equation}
According to Fermi's golden rule~\cite{dirac,fermi}, the decay probability $P$ is 
proportional to the level density at the final energy $\rho(E_i-E_{\gamma})$. 
The decay probability is also proportional to the squared transition matrix element
$|\langle f| \hat{T}|i \rangle|^2$ between initial $|i\rangle$ and final $\langle f|$ states, 
which is represented by
the $\gamma$-ray transmission coefficient ${\cal{T}}$ when averaged over many transitions
with the same transition energy $E_\gamma$. For now, let us assume
that the transmission coefficient depends only on $E_{\gamma}$, in accordance with the
gBA hypothesis.

The factorization given in Eq.~(\ref{eq:rhoT}) allows us
to simultaneously extract the functions $\rho$
and ${\cal{T}}$ from the two-dimensional probability landscape $P$.
The technique used, the Oslo method~\cite{Schiller00}, requires no
models for these functions.
In the present case~\cite{238Np},
we have fitted each vector element of the two functions to the following region of the $P$ matrix:
$3.0 \leq E_i \leq 5.4$~MeV and $E_{\gamma}> 0.84$~MeV.
The justification
of this standard procedure has been experimentally tested for many nuclei by the Oslo group~\cite{compilation}
and a survey of possible errors for the Oslo method was presented in Ref.~\cite{Lars11}.

The applicability of Eq.~(\ref{eq:rhoT}) and the quality of the least $\chi ^2$ fitting
procedure are demonstrated in Fig.~\ref{fig:doesitfit}.
The agreement is very satisfactory with a $\chi ^2_{\rm reduced} = 0.81$, and indicates
that the determination of the level density $\rho$
and the transmission coefficient ${\cal{T}}$ works well. In the following we assume that $\rho$ and ${\cal{T}}$ are
normalized according to the procedure in Ref.~\cite{Schiller00}. Thus, we introduce
a normalization factor $N$ in Eq.~(\ref{eq:rhoT}), which only depends on the initial excitation energy,
and rewrite:
\begin{equation}
N(E_i)P(E_{\gamma},E_i ) =  \rho(E_i-E_{\gamma}){\cal{T}}(E_{\gamma}) ,\
\label{eq:rhoTnorm}
\end{equation}
which determines the normalization factor by
\begin{equation}
N(E_i)=\frac{\int_0^{E_i}  {\cal T}(E_{\gamma}) \rho(E_i-E_{\gamma})\, {\mathrm{d}} E_{\gamma }
}{\int_0^{E_i} \, P(E_{\gamma},E_i)\, {\mathrm{d}} E_{\gamma}}.
\label{eq:nei}
\end{equation}

It is an open question
if the transmission coefficient ${\cal{T}}$ actually changes with excitation energy,
as this procedure gives an \textit{average} ${\cal{T}}$ for a rather wide range of initial
excitation energies in the standard Oslo method.
Hence, it is possible that variations of ${\cal{T}}$ as function
of initial (or final) excitation energy might be masked.
In the following, we will investigate whether ${\cal{T}}$ depends on initial
and final excitation energies in order to test the gBA hypothesis. For this we
collect $\gamma$ transitions from certain initial states or $\gamma$ transitions into
certain final states as illustrated in Fig.~\ref{fig:method}.

The idea is that, as the level density $\rho$ is known, the
$\gamma$ transmission coefficient can be studied in detail
per excitation energy bin simply by $NP/\rho$ from Eq.~(\ref{eq:rhoTnorm}).
\begin{figure*}[]
\begin{center}
\includegraphics[clip,width=1.9\columnwidth]{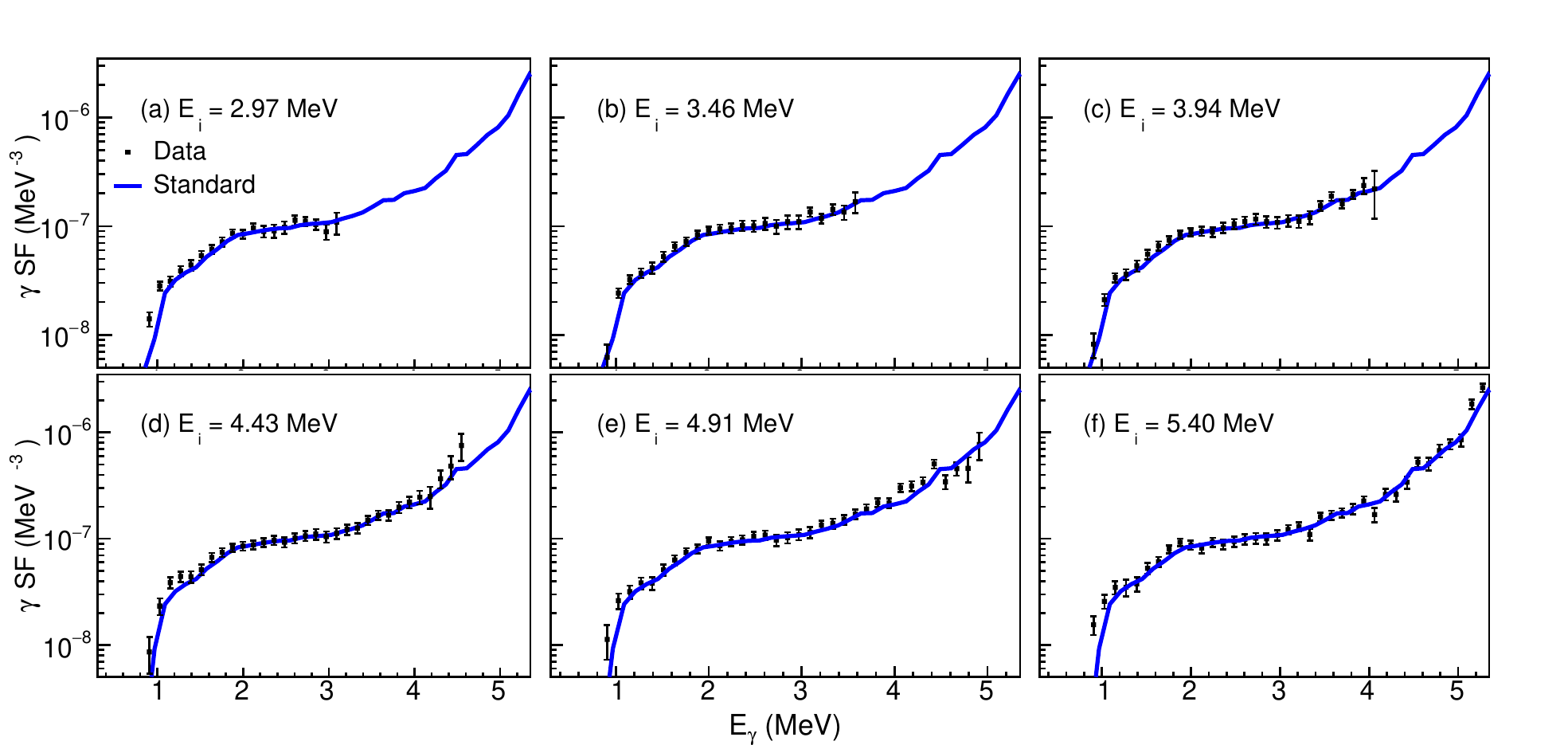}
\caption{(Color online) The $\gamma$SFs as function of initial excitation energies (data points),
see Eq.~(\ref{eq:Ei}). The blue curve
is obtained by the standard Oslo method, see Eq.~(\ref{eq:rhoT}). The excitation energy bins are 121 keV broad.}
\label{fig:initial}
\end{center}
\end{figure*}
\begin{figure*}[]
\begin{center}
\includegraphics[clip,width=1.9\columnwidth]{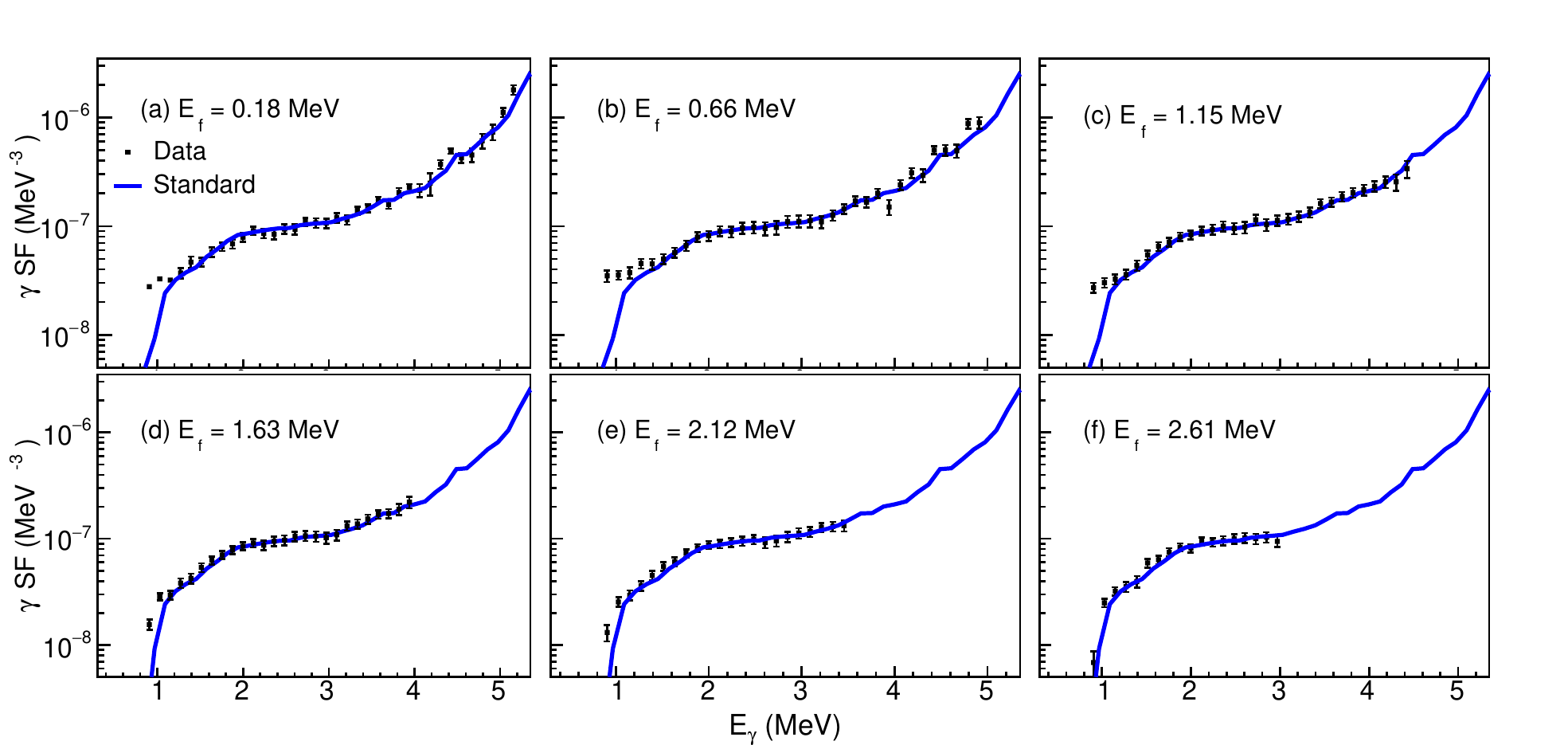}
\caption{(Color online) The $\gamma$SFs as function of final excitation energies (data points),
see Eq.~(\ref{eq:Ef}). See text of Fig.~\ref{fig:initial}.}
\label{fig:final}
\end{center}
\end{figure*}
More specifically, we get for initial states:
\begin{equation}
{\cal T} (E_{\gamma}, E_i ) =
\frac{N(E_i)P(E_{\gamma},E_i)}{\rho(E_i - E_{\gamma})},
\label{eq:Ei}
\end{equation}
or alternatively for final states:
\begin{equation}
{\cal T} (E_{\gamma},E_f) =
\frac{N(E_f + E_{\gamma})P(E_{\gamma},E_f + E_{\gamma})}{\rho(E_f)},
\label{eq:Ef}
\end{equation}
where $E_f=E_i-E_{\gamma}$.
One should note that
the normalization factor $N$ is calculated from the assumption that
both ${\cal T} (E_{\gamma},E_i)$ and ${\cal T} (E_{\gamma},E_f)$
fluctuate on the average around the excitation-independent ${\cal T} ( E_{\gamma})$,
see Eq.~(\ref{eq:nei}).

We now translate the $\gamma$ transmission coefficient into $\gamma$SF by~\cite{kopecky1990}
\begin{equation}
f(E_{\gamma}) = \frac{1}{2\pi}\frac{{\cal T}(E_{\gamma})}{E_{\gamma}^{2L +1}},
\label{eq:gSF}
\end{equation}
where we assume that dipole radiation ($L=1 $) dominates the $\gamma$ decay in the quasi-continuum.
This is motivated by known discrete $\gamma$-ray
transitions from neutron capture~\cite{kopecky1990}
and angular distributions of primary $\gamma$-rays measured at
high excitation energies~\cite{larsen2013}.

To check that the normalization function $N(E_i)$ is reasonable, we compare
the three $\gamma$SFs obtained from Eqs.~(\ref{eq:rhoT}), (\ref{eq:Ei}) and  (\ref{eq:Ef}),
where $f(E_{\gamma},E_i )$ and $f(E_{\gamma},E_f)$ are averaged over
initial and final excitation energies by
\begin{eqnarray}
f_i (E_{\gamma}) &=& \frac{1}{S_n - E_{\gamma}}\int_{0}^{S_n-E_{\gamma}} f (E_{\gamma},E_i )\, \mathrm{d}E_i,\\
f_f (E_{\gamma}) &=& \frac{1}{S_n - E_{\gamma}}\int_{E_{\gamma}}^{S_n} f (E_{\gamma},E_f)\, \mathrm{d}E_f,
\label{eq:3gsf}
\end{eqnarray}
respectively. Figure~\ref{fig:strength3} demonstrates that the three extraction methods
give $f(E_{\gamma}) = f_i(E_{\gamma})= f_f(E_{\gamma})$ within the experimental errors.
This supports the normalization function $N$ used in Eqs.~(\ref{eq:Ei}) and  (\ref{eq:Ef}).

We are now ready to compare our data with the gBA hypothesis. Figure ~\ref{fig:initial}
shows the initial excitation energy dependent $f(E_{\gamma}, E_i)$ compared to $f(E_{\gamma}$) obtained
with the standard Oslo method (blue curve). The excitation-energy bins are $\Delta E_i = 121$~keV wide,
and only every fourth gate is shown. The overall agreement is excellent, and
the same holds also for all the bins not shown. It is clear that each of these $\gamma$SFs are
built on a specific initial excitation-energy gate, but with no specific final state, as illustrated
in Fig.~\ref{fig:method}. However, for a given
 $E_{\gamma}$ and $E_i$, the final excitation energy is determined. Since all
data points coincide with $f(E_{\gamma})$,
this also indicates independence of the final state.

Any potential dependence of the final state is best analyzed by $f(E_{\gamma},E_f )$ as given
by Eq.~(\ref{eq:Ef}) and shown in Fig.~\ref{fig:final}.
Again, we find an excellent agreement between the various $\gamma$SFs with $\gamma$ transitions
into specific final excitation-energy bins. However, there are discrepancies for $E_{\gamma} < 1$~MeV, which feed
final states below $\approx 1$~MeV. At these energies, $f(E_{\gamma},E_f)$ shows an increase compared to the average
$f(E_{\gamma})$. These $\gamma$ transitions could possibly be due to vibrational modes built on
the ground state, and, if this be true, not part of a general $\gamma$SF extracted
in the quasi-continuum with the standard Oslo method. Vibrational levels are
strongly populated in inelastic scattering, such as the reactions $^{237,239}$Np($d,d^{\prime}$)$^{237,239}$Np
performed by Thompson {\em et al.}~\cite{dd}. In that work, levels built on vibration modes were seen
for excitation energies in the $\approx 0.9$-MeV and $\approx 1.6$-MeV regions.
A similar population of vibrational states
has been observed in the $^{238}$U($^{16}$O,$^{16}$O$^{\prime}$)$^{238}$U and $^{238}$U($\alpha$,$\alpha^{\prime}$)$^{238}$U
reactions~\cite{aa}. By means of $\alpha\gamma$-coincidences, a concentration
of $E_{\gamma}\approx 1$~MeV transitions depopulating $\beta$-, $\gamma$- and octupole vibrational bands 
has been seen.
Thus, the enhanced $\gamma$SF found in our data at low excitation energies with $E_{\gamma} \approx 1$~MeV
is likely due to deexcitation of vibrational structures, which do not show up in the
high level density quasi-continuum.

The excellent agreement between excitation energy dependent
and independent $\gamma$SFs indicate that PT
fluctuations are small compared to the experimental errors for the system studied.
For the $\chi_{\nu}^2$ distribution, which governs the PT fluctuations,
the relative fluctuations are given by $r=\sqrt{(2/\nu)}$ where
$\nu$ is the number of degrees of freedom~\cite{murray1975}. Typically, we have at 2.0 and 4.0
MeV of excitation energies, $\approx 1200$ and $\approx 120000$ levels within the 121-keV excitation energy bins.
Taking the number of levels as the number of degrees of freedom,
we obtain $r =  4.1$~\% and 0.4~\%, respectively, which
should be compared with the data error bars of Figs.~\ref{fig:initial} and \ref{fig:final} of
typically 10\%. Thus, in the $^{238}$Np case, the PT
fluctuations are smaller than the statistical errors and
not significant. However, for systems with less than $\approx 200$ levels
per bin the PT fluctuations will exceed the experimental statistical error of 10\%.
This gives guidance for the necessary statistics and the required level density
for the gBA hypothesis to be fulfilled.

In conclusion, we have studied the $\gamma$-ray strength function between
well defined excitation energy bins in $^{238}$Np.
For the first time, the generalized Brink-Axel hypothesis has been
verified in the nuclear quasi-continuum. The
discrepancies seen in the low excitation energy region are probably caused by
decay of vibrational states built on the ground state. These excitation modes
are not part of the $\gamma$-ray strength function of the
quasi-continuum. The validity of the generalized Brink-Axel hypothesis
requires that Porter-Thomas fluctuations are low by averaging over a sufficient amount of levels
compared to the experimental errors.

\acknowledgements
The authors wish to thank J.C.~M{\"{u}}ller, E.A. Olsen, A. Semchenkov and J. Wikne
at the Oslo Cyclotron Laboratory for providing excellent experimental conditions.
A.C.L. acknowledges financial support through the ERC-STG-2014 under grant agreement no. 637686.
S.S. and G.M.T. acknowledge financial support by the NFR under project grants no. 210007 and 222287, respectively.

\end{document}